# Resonance Raman mapping as an interface phonon probe in Si-SiO$_2$ nanocomposites

**Ekta Rani[1], Alka Ingale[1*], A. chaturvedi[2], M. P. Joshi[2] and L. M. Kukreja[2]**

[1]Laser Physics Applications Section, Raja Ramanna Centre For advanced Technology, Indore-452013, INDIA.

[2]Laser Material Processing Division, Raja Ramanna Centre For advanced Technology, Indore-452013, INDIA.

[*]E-mail: alka@rrcat.gov.in

Intermediate frequency range (511 - 514 cm$^{-1}$) Si phonons in Si-SiO$_2$ nanocomposites are shown to have contribution from both core[1] and surface/interface[1] Si phonons, where, ratio of contribution of the two depends on the size of a Si nanocrystal. Further, laser heating experiment shows that contribution of the core phonon increases due to increase in size of a nanocrystal. Wavelength dependent Raman mapping reveals that interface phonons are observable due to Resonance Raman scattering. This can well be corroborated with the absorption spectra. This understanding can be gainfully used to manipulate and characterize Si-SiO$_2$ nanocomposite, simultaneously for photovoltaic device applications.

Si nanocrystals (NCs) mainly embedded in dielectric matrix material have received great attention in the past few decades due to their importance in fundamental physics and potential applications of Si NCs in third generation photovoltaics,[2] optoelectronic devices,[3] light emitting diodes[4]. To overcome Shockley-Quisser limit for the efficiency of single band gap materials, third generation photovoltaic devices employ multiple optical gaps in a single device. In this regard, interface of Si-SiO$_2$ nanocomposite (NCp) is of intense current interest and has been specifically well studied theoretically.[5-6] Presence of oxygen at the surface is expected to play an important role in affecting the electronic and optical properties of Si NCs, specially for smaller sizes. The embedded interface in case of Si-SiO$_2$ is difficult to study directly through experiments, however indirect studies using PL spectroscopy,[7] X-ray photoemission spectroscopy[8] etc. has been attempted. In this paper, we present Raman spectroscopy as a research tool to investigate Si-SiO$_2$ interface. Raman spectroscopy is found to be sensitive to size and interface in case of Si-SiO$_2$ NCp.[1,9] We report herein, the use of Raman mapping and spectroscopy as a local probe to exploit this sensitivity, to obtain further information on the interface of Si NCs in SiO$_2$ matrix with large size distribution.

Si-SiO$_2$ multilayer NCp is grown on sapphire by ablation of Si and SiO$_2$ targets alternatively in pulsed lased deposition chamber by varying the deposition time of silicon as 1) E1 : 45 s, 2) E2 : 90 s, 3) E3 : 120 s, 4) E4 : 150 s, 5) E5 : 180s, 6) E6 : 210 s and deposition time for SiO$_2$ layer is kept constant for all the samples. Other growth and sample details are given in our previously published papers.[1,10] Raman spectroscopic/mapping measurements are performed in backscattering geometry at room temperature using Acton 2500i (single) monochromator with air cooled CCD detector, a part of scanning probe microscopy_integrated Raman system set up, WiTec; Germany (spectral resolution ~ 2.5 cm$^{-1}$). Absorption spectroscopy measurements are performed using CARY 50 spectrophotometer.

Raman mapping shows phonon frequency varying from 495 - 519 cm$^{-1}$ in, Si-SiO$_2$ NCps, wherein these NCps are made up of clusters of Si NCs in SiO$_2$ matrix.[1] Further, we have attributed phonon frequencies observed in the range 515 - 519 cm$^{-1}$ (HF) and 495 - 510 cm$^{-1}$ (LF) to core and interface of Si NCs in Si-SiO$_2$ NCps, respectively.[1] In the present paper, we investigate the origin of Si phonon frequencies observed in the intermediate (IF) range 511 - 514 $cm^{-1}$. Raman mapping measurements are performed at many sites in all the samples for IF phonons. Raman spectra for IF phonons show mainly two types of asymmetric line shapes : i) IF1 with counts/s (C/s) in the range 400 - 600 C/s and frequency ~ 511 $cm^{-1}$ (Figure 1a) , ii) IF2 with C/s in the range 50 - 200 C/s and frequency ~ 513 $cm^{-1}$ (Figure 1b) with FWHM ~ 7.5 - 9 $cm^{-1}$. Further, this asymmetric line shape is quite different than that of HF phonons fitted using phonon confinement model (PCM).[12]

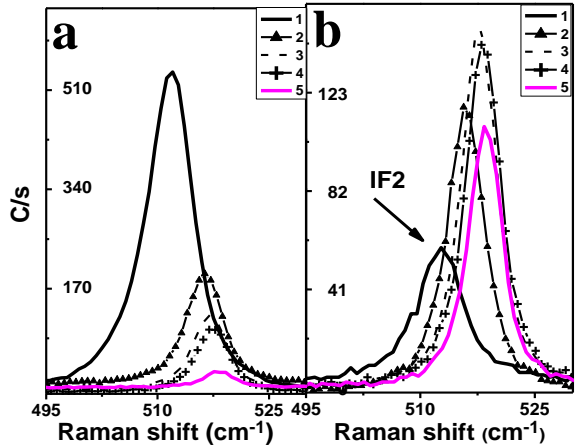

Figure 1: Change in a) IF1 phonon and b) IF2 phonon during 5 steps of LHC experiment performed using 441.6 nm excitation with power density ~ 5 $kW/cm^2$. LHC steps are noted in the text.

We find that an asymmetric line shape of IF phonons can be fitted using superposition of LF and HF phonons. The consideration here is that the intermediate size NCs can have contributions from both interface and core phonon with a understanding that smaller sizes (< 40 Å) and larger size (> 60 Å) give rise to LF and HF phonons, respectively.[10,13-14] The laser heating and cooling (LHC) experiment[1] is performed on the desired IF phonon to confirm this interpretation. LHC steps are: 1st step; 2nd step: *laser on for 16 mins*; 3rd step : *Laser off for 40 mins*; 4th step

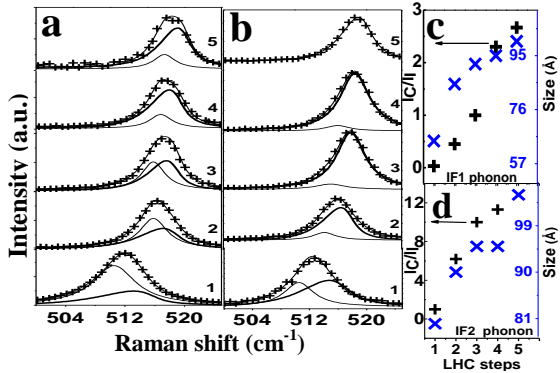

Figure 2: Change in a) IF1 phonon, b) IF2 during LHC experiment. c) & d) Change in ratio of intensity of HF to LF phonon and size for these two phonons during LHC experiment, respectively. LHC steps are noted in the text.

: *Again laser on for 16 mins*; 5th step : *Laser off for 3 hrs*. Raman spectrum is recorded at the end of each step. We observe a pattern similar to that of LF phonon i.e. blue shift up to ~ 518 $cm^{-1}$ with decrease in FWHM (Figure 1a & b). However, there is one major difference in the behavior of these two IF phonons, i.e. IF1 phonons show huge decrease in the intensity (Figure 1a) whereas, IF2 phonons show increase in the intensity during LHC (Figure 1b). Line shapes of IF phonons are fitted with a Lorentzian and a PCM, accounting for contribution of LF (interface) and HF (core) phonons, respectively (Figure 2). The variation of size and ratio of the intensity of HF and LF phonon for both IF1 and IF2 phonons is plotted in Figure 2 c & d. This shows that in case of both IF1 and IF2 phonons, there is increase in size of Si NC and thereby increase in the contribution of the core phonon. At the end of 5$^{th}$ step of LHC experiment, IF1 phonon cannot be fitted using PCM, whereas IF2 can be satisfactorily fitted using PCM (Figure 2). This suggests that LHC experiment has led to more Si-Si like environment at the surface and thus increase in size of NC.

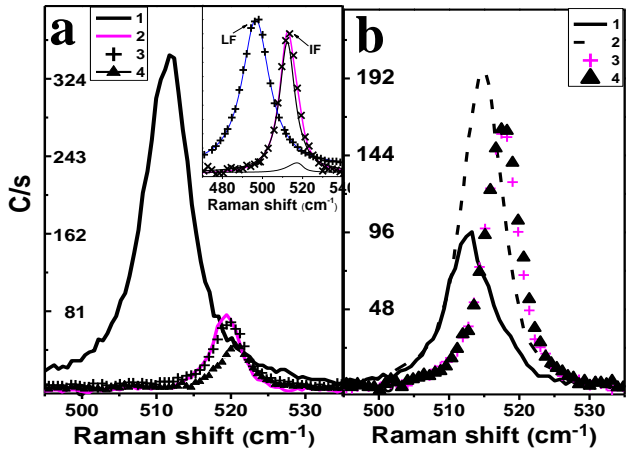

Figure 3: Change in a) IF1 phonon and b) IF2 phonon during 1) 1$^{st}$ step of LHC, 2) 3$^{rd}$ step of LHC, 3) 5$^{th}$ step of LHC and 4) 2 hrs heating. Inset shows normalized Raman spectra showing conversion of LF (shifted in y-axis for clarity) to IF phonon on continuous laser heating.

To get a clue of the huge change observed in the intensity of these two types of IF phonons during LHC experiment, we decided to observe the effect of continuous heating for 2hrs after the LHC steps. During this IF2 phonon is not showing any change in frequency, FWHM and intensity (Figure 3b). This behavior is similar to HF phonons[1] and confirms our interpretation that this is a core phonon. Here also IF1 phonons show blue shift (Figure 3a). The Si phonon frequency observed at 520.8 cm$^{-1}$ shows significant asymmetry (PCM: 100 Å) and smaller contribution of a Lorentzian. A lineshape and C/s of this phonon is similar to IF2 phonon. Both these frequencies seem to be blue shifted and are indicative of generation of compressive stress due to this heating process.

Local temperature is calculated using interface (Stokes, anti-Stokes Raman data) comes out to be incorrect (varying from ~ 550 K to 140 K on laser heating) similar to the observation of LF phonons in our earlier work,[1] suggesting Resonance Raman scattering for IF phonons too. This is further studied, using wavelength dependent Raman mapping/spectroscopy. Representative Raman mapping performed with similar power density ~ 4-5 kW/cm$^2$ (0.5 s acquisition time) using 441.6 nm (Figure 4b), 488 nm (Figure 4c) and 514.5 nm (Figure 4d) excitation for sample E4 is shown in Figure 4. Raman image is created using LF, IF and HF phonons. Corresponding Raman spectra is shown in Figure 4f-4h showing maximum intensity obtained for LF, IF and HF phonon for these three lasers lines. Raman images show reduction in cluster size as the excitation source changes from 441.6 to 514.5 nm. This will be discussed later in the text. It is noteworthy to observe that LF and IF phonons could be observed only using 441.6 and 488 nm excitations, whereas, HF phonons are observed with all three excitations. Further, the ratio of intensities of interface and core phonons ($I_{LF}/I_{HF}$) is > 1. This is quite surprising knowing that these phonons are coming from much smaller volume (interface ~ 8-10 Å[15]). This suggests that either there is very large density of very small Si NCs, where from interface phonon contribution is dominant or these phonons are observable due to Resonance Raman scattering. If it is the first case, interface phonons should then be observable with 514.5 nm excitation source also. Excitation dependence of LF and IF phonon's intensity clearly shows that enhancement of Raman signal due to resonance is crucial to the observance of these phonons. It is important to note that Miska et al.[16] have also observed resonance Raman scattering for phonon frequency ~ 490 cm$^{-1}$ in Si-SiO$_2$ NCp. Further, it is interesting to note that HF phonons show non resonant behavior like bulk Si phonon and thus these phonons are observable with all the three excitation laser sources with similar intensity. This is also consistent with the fact that HF phonons give correct temperature from Stoke/anti-Stoke Raman data.[1]

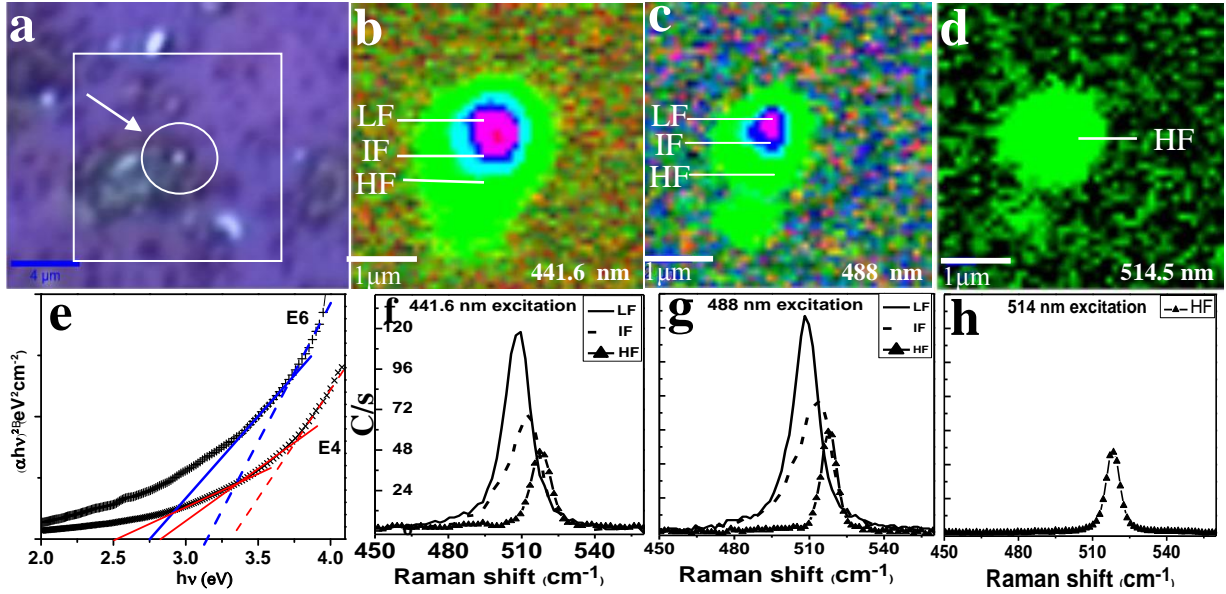

Figure 4: a) Optical image of sample E4, corresponding Raman image (step size ~ 0.15 µm) of marked region with excitation as b) 441.6 nm , c) 488 nm , d) 514 nm showing absence of LF and IF phonons with 514 nm as excitation laser, e) (αhv)$^2$ vs hv graph for sample E4 and E6 obtained from absorption spectra measured at Brewster angle and 4f-4h) compares relative intensity of LF and IF to HF phonons for wavelengths noted in the figure. C/s for all the three Raman spectra is shown in Figure 4f. Spectral and spatial resolution varies from ~ 4 to 3 cm$^{-1}$ and from ~ 0.98 to 1.14 µm for 442 - 514 nm, respectively.

It is interesting to note further that in the sample E6, LF and IF phonons are observable only with 441.6 nm laser excitation. To understand this observation, we have performed absorption spectroscopy measurements on these samples. However, due to presence of interference fringes in the absorption spectra, correlation with resonance Raman scattering is not feasible. The interference fringes free absorption data is then obtained by measuring the absorption at Brewster angle (instead of normal incidence) using p-polarized light to remove reflected component from the substrate/sample interface.[17] This absorption spectra shows absorption edges ~ 2.50 eV, 2.9 eV and 3.1 eV (Table 1) for all the samples except E6 where absorption edges ~ 2.3, 2.75 eV and 3.1 eV are observed (Figure 4e, Table 1). This clearly brings out proximity of real energy levels to the laser excitations and corresponding Raman signal enhancement due to resonance. This also explains observance of LF and IF phonons, only for 441.6 nm (2.8 eV) in case of E6. The results for all samples are summarized in Table 1. For sample E4, $I_{LF}/I_{HF}$ is highest for both 441.6 nm and 488 nm excitation source as compared to other samples. This can also be further correlated with observed absorption edges ~ 2.55 eV, 2.8 eV (Table 1) for this sample, which almost exactly matches with excitation wavelengths 448 nm (2.54eV) and 441.6 nm (2.8 eV) respectively. Absorption edge ~ 3.1 eV corresponds to direct band gap of Si, whereas other absorption edges ~ 2.5 eV and 2.9 eV are related to interface of Si-$SiO_2$ NCp originating from smaller size NCs.[7] Further, our observation of these band edges are also in agreement with recent eliipsometric study on Si-$SiO_2$ NCp, which also show band edges ~ 2.8 eV and 2.5 eV for smaller size Si NCs (20 - 40 Å).[18]

Table 1: Ratio of intensity of LF to HF phonon for all samples with 441.6 nm and 488 nm as excitation sources. Multiple absorption edges calculated from $(\alpha h\nu)^2$ vs $h\nu$ graph (Figure 4e) in the range of interest (350 - 520 nm) for all samples.

| Sample | $I_{(LF)}/I_{(HF)}$ (441.6 nm) | $I_{(LF)}/I_{(HF)}$ (488 nm) | Absorption edges (eV) |
|---|---|---|---|
| E1 (45s) | 1.8 | 1.5 | 2.5, 2.88, 3.2 |
| E2(90s) | 2.0 | 1.4 | 2.5, 2.9, 3.1 |
| E3(120s) | 1.8 | 1.2 | 2.48, 2.9, 3.2 |
| E4(150s) | 3.0 | 2.7 | 2.55, 2.8, 3.3 |
| E5(180s) | 2.1 | 1.8 | 2.6, 2.92, 3.1 |
| E6(210s) | 1.5 | No LF | 2.75, 3.1 |

We further performed wavelength dependent Raman mapping on Si-Al$_2$O$_3$ (amorphous Al$_2$O$_3$) NCp film grown on sapphire. Phonon frequencies in the range 490 - 519 cm$^{-1}$ are observed only with 441.6 and 488 nm excitation. However, only HF phonons (515 - 519 cm$^{-1}$) are observed all the three excitations, including 514.5 nm excitation. This result corroborates well with the absorption spectrum of the same.

Wavelength dependent Raman mapping (Figure 4) shows that i) cluster size reduces as the wavelength changes from 441.6 nm to 514.5 nm and ii) with 514.5 nm excitation, HF phonons are observed also in the region, where LF and IF phonons are observed with 488 and 441.6 nm excitations. To understand these observations, Raman mapping is performed with varying the focus along the depth. The results suggest that HF phonons are coming from bottom of the cluster, whereas IF and LF phonons are coming from central and top region of a cluster, respectively. Highest depth of focus for 514.5 nm excitation and non resonant nature of HF phonons explains the presence of HF phonons at this excitation, from the region, where LF and IF phonons are observed at 488 and 441.6 nm excitations. The decrease in overall cluster size from 441.6 to 514.5 nm is found to be due to lower scattering efficiency at 514.5 nm leading to observable signal only coming from larger scattering volume region. Details will be published elsewhere.[11]

**Understanding the origin of two different phonons in the frequency range 511 - 514 cm$^{-1}$:** Results of LHC experiment and continuous laser heating for longer time can be explained considering that these two IF1 and IF2 phonons come from two different intermediate size NCs, $x_1$ and $x_2$, where $x_1 < x_2$. For smaller size $x_1$, interface phonon contribution is dominating (Figure 2a) whereas, for larger size $x_2$, core and interface phonon contribution is comparable (Figure 2b). The decrease in the intensity of IF1 phonon on heating can be correlated as due to moving away from the resonance condition for LF phonons due to increase in size and thereby increase in core phonon contribution. Increase in intensity of IF2 phonon on heating can be said to be due to increase in the size of Si NC i.e. scattering volume. The non resonant nature of core phonon further explains the lesser C/s for IF2 phonon as compared IF1 phonon in the beginning of the LHC experiment (Figure 1).

It is found that during the LHC experiment, more and more Si-O bonds are getting converted to Si-Si like environment causing increase in size of Si NC. This increases the intensity of IF2 phonons, which already has comparable core phonon contribution. It is important to note that C/s of IF1 phonons at the end of LHC experiment is close to C/s of IF2 phonons at the beginning of the experiment. This further confirms the increase in size $x_1$ and it is approaching size $x_2$ on LHC. This clearly suggests conversion of IF1 phonon to IF2 phonon and IF2 phonons to HF phonons. However, LF phonons did not show conversion of LF phonon to IF phonons during the LHC experiment.[1] This is looked into by continuous laser heating for longer time of 3 hrs and 30 mins on LF phonons. It is observed that LF phonon (~ 497 cm$^{-1}$: Lorentzian line shape) converted to IF phonon ~512.9 cm$^{-1}$ (Inset of figure 3a) after continuous heating. The asymmetric lineshape of this phonon can be fitted with superposition of LF and HF phonon. This shows conversion of LF to IF phonons and thus further confirming our interpretation.

It may be interesting to see the correlation of size of NC, wavelength dependent Raman spectra with different interface or passivation of Si NCs reported in literature. Hydrogen passivated Si NCs of sizes ~ 20 - 60 Å excited with 514.5 nm and 496.5 nm show HF[19, 20] and LF[21] phonon, respectively. Raman spectra of Si-SiO$_2$ NCp (size ~ 30 - 50 Å) excited with mainly 514.5 nm laser show mainly observance of core phonons.[22,23] However, Si NCs of size ~ 20 - 40

Å excited using 488 nm, shows presence of Si phonon frequency ~ 495 cm$^{-1}$ and 517 cm$^{-1}$ simultaneously. Further, peak ~ 495 cm$^{-1}$ disappear for Si NCs of size ~ 63 Å.[13] These observations are consistent with our interpretation of LHC experiment as noted above. Further, in some cases, Raman spectra of Si-SiO$_2$ NCp (size of NC ~ 30 - 60 Å) excited with 488 nm laser only show presence of core phonons.[24] This data further supports that occurrence of surface/interface phonons depends on two factors 1) size of Si NC and 2) excitation wavelength. This in turn depends on matrix/passivation of Si NCs, as it leads to different electronic band structure.[1] The above mentioned and our data together suggests that LF, IF and HF phonons are observed from sizes < 40 Å, 40 Å to 60 Å and > 60 Å, respectively.

In our earlier work, we have shown using laser heating and cooling experiment (LHC) that Si phonon frequencies in the range LF and HF phonons are attributed to originate from interface (smaller NCs) and core (larger NCs) of Si NC in Si-SiO$_2$ NCp, respectively.[1] In this paper, Raman spectroscopy monitored Laser heating cooling experiment shows that phonon frequency in the intermediate range 511 - 514 cm$^{-1}$ is originating from intermediate size NCs where, both surface and core of Si NC contribute simultaneously. Blue shift and increase and decrease in the intensity for two types of interface phonons IF1 and IF2 can both be correlated to increase in size of Si NC and thereby leading to a dominant contribution of core phonons. Further, temperature obtained from Stoke-AntiStokes Raman spectroscopy and results of wavelength dependent Raman mapping in corroboration with absorption measurements show that resonance Raman scattering is crucial for the observance of interface phonons. This also further explains large Raman signal observed from Si-SiO$_2$ interface and the Raman data variation observed in the literature for Si phonon frequency in Si-SiO$_2$ NCps, Si NCs passivated using hydrogen etc. This understanding can be advantageously used to obtain information of the interface and estimation of size in Si-SiO$_2$ nanocomposites nondestructively and without needing any sample preparation. This can be very useful for characterization of a fabricated device. It is also important to note that on laser heating at low power, we found that if enough time is given, then Si NCs grow in size to give larger core phonon contribution i.e. LF phonon gets converted to IF phonon and IF phonons gets converted to HF phonon confirming our attributions of these phonons. This can further allow Raman spectroscopy monitored, controlled manipulation of the device properties using laser.

We acknowledge continual support of Dr. H. S. Rawat for this work. We also acknowledge Dr. C. Mukherjee and K. Rajiv for the help provided in obtaining the absorption data. Ms Ekta Rani wishes to acknowledge "Homi Bhabha National Institute", India for providing research fellowship during the course of this work.